\documentstyle[fleqn,12pt]{article}
\newcommand{\eeq}{\end{eqnarray}}
\newcommand{\beq}{\begin{eqnarray}}
\newcommand{\bD}{{\mbox{\boldmath$D$}}}
\newcommand{\bE}{{\mbox{\boldmath$E$}}}
\newcommand{\bB}{{\mbox{\boldmath$B$}}}
\newcommand{\bH}{{\mbox{\boldmath$H$}}}
\newcommand{\bF}{{\mbox{\boldmath$F$}}}
\newcommand{\bpi}{{\mbox{\boldmath$\pi$}}}
\newcommand{\bp}{{\mbox{\boldmath$p$}}}
\newcommand{\bv}{{\mbox{\boldmath$v$}}}
\newcommand{\ba}{{\mbox{\boldmath$a$}}}
\newcommand{\bq}{{\mbox{\boldmath$q$}}}
\newcommand{\bx}{{\mbox{\boldmath$x$}}}
\newcommand{\br}{{\mbox{\boldmath$r$}}}
\newcommand{\bN}{{\mbox{\boldmath$N$}}}
\newcommand{\om}{{\mbox{\boldmath$\omega$}}}
\newcommand{\bP}{{\mbox{\boldmath${\cal P}$}}}
\newcommand{\bR}{{\mbox{\boldmath${\cal R}$}}}
\newcommand{\bS}{{\mbox{\boldmath${\cal S}$}}}

\newtheorem{TH}{Theorem}
\date{}
\title{Canonical formalism for the Born-Infeld particle}
\author{\\ \\ Dariusz Chru\'sci\'nski\footnotemark \\
       Fakult\"at f\"ur Physik, Universit\"at Freiburg\\
       Hermann-Herder-Str. 3, D-79104 Freiburg, Germany}

\begin{document}

\def\thefootnote{\relax}\footnotetext{$^*$On leave from
 Institute of Physics, Nicholas Copernicus
University, ul. Grudzi\c{a}dzka 5/7, 87-100 Toru\'n, Poland.}

\maketitle

\begin{abstract}

In the previous paper [1] it was shown  that the nonlinear Born-Infeld field 
equations supplemented
by the ``dynamical condition'' (certain boundary condition for the
field along the particle's trajectory) define perfectly deterministic
 theory, i.e. particle's trajectory is determined without any equations 
of motion. In the present paper we show that this theory possesses
mathematically consistent Lagrangian and Hamiltonian formulations.
Moreover, it turns out that the "dynamical condition" is already present
in the definition of the physical phase space and, therefore, it is
a basic element in the above theory.

\end{abstract}

\vspace{4cm}

Freiburg THEP--98/5

\newpage

\section{Introduction}

  Born-Infeld electrodynamics
\cite{BI} was proposed in the thierties as an alternative
for Maxwell theory (see also \cite{IBB} for a useful review).
The main motivation of Born and Infeld was to construct a theory which
has classical solutions representing electrically charged particles
with finite self-energy. Born-Infeld theory indeed admits  such  solutions
(recently Gibbons proposed \cite{BIons} to call them "BIons").
 However,  after  Dirac's paper \cite{Dirac}
on a classical electron and the birth of quantum electrodynamics in
the forties Born-Infeld theory was almost totally forgoten for a long time.

Recently, there is a new interest in this theory
due to  investigations in  string theory. It turns out that 
some very natural objects in this theory, so called  D-branes,  are
 described by  a kind of nonlinear Born-Infeld
action (see e.g. 
\cite{BIons} and \cite{brane}). Moreover, due to the remarkable
interest in field and string theory dualities \cite{Olive}, the duality
invariance of Born-Infeld electrodynamics was studied in great details
\cite{duality} (actually this invariance was already observed 
by Schr\"odinger \cite{Schrodinger}).

Our motivation to study Born-Infeld electrodynamics 
is not a string theory but a dynamics of a classical point charge. 
There are following reasons to consider nonlinear electromagnetism:
  it is well known that Maxwell 
electrodynamics when applied to point-like objects is inconsistent
(see  \cite{Rohrlich} for the review). This
inconsistency originates in the infinite self-energy of the point charge.
In the Born-Infeld theory this self-energy is already finite. 
Therefore, one may hope that in the theory which gives finite value 
of this quantity it would be possible to describe the particle's
self-interaction in a consistent way. Moreover, the assumption, that
the theory is effectively nonlinear in the vicinity of the charged
particle is very natural from the physical point of view and
this we already  learned from quantum electrodynamics (Born tried
to make contact with quantum field theory by identifying Born-Infeld
Lagrangian as an effective Euler-Heisenberg Lagrangian \cite{Euler}. It
has been shown \cite{Hagiwara1} that the effective Lagrangian can coincide
with those of Born and Infeld up to six-photon interaction terms only).

 We consider 
very specific model of nonlinear theory because, among other nonlinear
theories of electromagnetism, Born-Infeld theory possesses very
distinguished physical properties \cite{Hagiwara2}. For example
it is the only causal spin-1 theory \cite{Plebanski1} (apart from 
the Maxwell one).  Recently, Born-Infeld electrodynamics was
successfully applied \cite{multipole} as a model for generation
of multipole moments of  charged particles.

In the previous paper \cite{charge} we studied the electrodynamics
of a point charge in the Born-Infeld nonlinear theory. 
Due to the nonlinearity of the field equations it is impossible
to derive separate equations of motion for the charged particle
corresponding  e.g.  to the celebrated Lorentz-Dirac equation in 
the Maxwell case.  Could we, therefore,
determine the particle's trajectory without equations of motion?
 We showed \cite{charge} that it is in fact possible. 
 Analysing the interaction between charged particle
and nonlinear electromagnetism we proved that the conservation of
total four-momentum of the composed (particle+field) system is
equivalent to the certain boundary condition for the Born-Infeld
field which has to be satisfied along the particle's trajectory.
We call it ``dynamical condition'' (formula (\ref{dynamical}) in the
present paper)
 because, roughly speaking, it replaces particle's equations of motion. 
Field equations supplemented by this condition define perfectly
deterministic theory, i.e. initial data for the particle and field 
uniquely  determine the evolution of the system. 

In the present paper we show that the theory derived in \cite{charge} 
possesses consistent Lagrangian and Hamiltonian structures. It is important
because any reasonable physical theory could be formulated this way.
Therefore, we expect that also Born-Infeld electrodynamics completed by
the dynamical condition (\ref{dynamical}) is no exception in this rule.
Moreover, we claim that the mathematically well defined canonical
structure is a necessary condition for the theory to be consistent.
It should be stressed that our model of a charged particle is different 
from that
of Born and Infeld, i.e. our particle is not a purely electromagnetical
"BIon". Nevertheless, we call it a "Born-Infeld particle".
A particle's mass, which appears e.g. in formula (\ref{Newton}),
could be interpreted as an effective mass, 
i.e. a "mechanical" mass completed by 
"radiative corrections" which are due to the electromagnetic interaction.
The finite self-energy of a charge (i.e. mass of its  Coulomb field)
 is already contained in the field energy.
Due to the nonlinearity of the theory there is no way to separate this
self-energy from the total field energy (in Maxwell theory this separation
enables one to perform the mass renormalization \cite{Rohrlich}).
 It turns out that the theory with a purely
electromagnetical charged particle (contrary to the one considered
in the present paper)  does not have any consistent
Hamiltonian formulation (this observation was made long ago by Pryce 
\cite{Pryce}).

The paper is organised as follows: in the next section we briefly sketch
main results of \cite{charge}. In section~3 a 2-nd order
particle's Lagrangian is constructed and it is proved that the corresponding
Euler-Lagrange equations are equivalent to our dynamical condition.
Then in section~4 we present a Hamiltonian structure
together with a Poisson bracket in section~5. Finally, we
make some conclusions and outline possible generalizations.

\section{Dynamical condition}

In the present section we briefly sketch the main result presented in
\cite{charge}.
The Born-Infeld nonlinear electrodynamics \cite{BI}  is based
on the following Lagrangian (we use the Heaviside-Lorentz system of
 units with the velocity of light $c=1$):
\begin{eqnarray}    \label{Lag-BI}
{\cal L}_{BI} :=  \sqrt{-\mbox{det}(b\eta_{\mu\nu})}
- \sqrt{-\mbox{det}(b\eta_{\mu\nu} + F_{\mu\nu})}  
\ = \ { b^2}\left(1- \sqrt{1-2b^{-2}S - b^{-4}P^2}\right)\ ,
\end{eqnarray}
where $\eta_{\mu\nu}$ denotes the Minkowski metric with the signature
$(-,+,+,+)$.
The standard Lorentz invariants
$S$ and $P$ are defined by:
$S = -\frac 14 F_{\mu\nu}F^{\mu\nu}$ and $P = -\frac 14
F_{\mu\nu}\tilde{F}^{\mu\nu}$ ($\tilde{F}^{\mu\nu}$ denotes the dual tensor).
The arbitrary parameter ``$b$'' has a dimension of a field strength (Born
and Infeld called it the {\it absolute field}) and it measures the
nonlinearity of the theory. In the limit $b \rightarrow \infty$ the
Lagrangian ${\cal L}_{BI}$ tends to the Maxwell Lagrangian $S$.

Adding to (\ref{Lag-BI}) the standard electromagnetic interaction term
``$j^\mu A_\mu$'' we may derive the inhomogeneous field equations
\beq    \label{field-eqs}
\partial_\mu G^{\mu\nu} = - j^\nu\ ,
\eeq
where $G^{\mu\nu} := -2
 {\partial {\cal L}_{BI}}/{\partial F_{\mu\nu}}$. Equations
(\ref{field-eqs}) have formally the same form as Maxwell equations. 
What makes the theory effectively nonlinear are the constitutive
relations, i.e. relations between inductions $(\bD,\bB)$ and
intensities $(\bE,\bH)$:
\beq
\bE(\bD,\bB) &=& \frac{1}{b^2R} \left[ (b^2 + \bB^2)\bD -
(\bD \bB)\bB \right] \ ,  \label{E}\\
\bH(\bD,\bB) &=& \frac{1}{b^2R} \left[ (b^2 + \bD^2)\bB -
(\bD \bB)\bD \right] \ ,  
\eeq
with $R:= \sqrt{1 + b^{-2}(\bD^2 +\bB^2) +  b^{-4}(\bD\times\bB)^2}$.

Now, let us assume that the external current $j^\mu$ in (\ref{field-eqs})
is produced by a point-like particle moving along the time-like trajectory
$\zeta$. Due to the nonlinearity, the system (\ref{field-eqs}) 
 is very complicated to analyse. In particular, contrary to the
Maxwell case, we do not know the general solution to the inhomogeneous
Born-Infeld field equations. The main idea of \cite{charge} (it was proposed
in the Maxwell case in \cite{KIJ}) was as follows: instead of solving 
distribution equations
(\ref{field-eqs}) on the entire Minkowski space-time $\cal M$ let us
treat them as a boundary problem in the region ${\cal M}_\zeta :=
{\cal M} - \{\zeta\}$, i.e. outside the trajectory. Obviously, in order to well 
pose the problem we have to find an appropriate boundary condition 
which has to be satisfied along $\zeta$, i.e. on the boundary 
${\cal M}_\zeta$. 

To find this boundary condition we have analysed the
asymptotic behaviour of the fields in the vicinity of a charged particle. 
The simplest way to do it is to use the particle's rest frame. Let $r$ denotes
the radial coordinate, i.e. a distance from a particle in its rest frame. Any
vector field ${\bF} = {\bF}({\br})$ may be formally expanded in the
vicinity of a charge:
\beq
\bF({\br}) &=& \sum_{n} r^n \bF_{(n)}\ ,
\eeq
where the vectors $\bF_{(n)}$  do not depend on $r$. The crucial observation
is that the most singular part of $\bD$ field behaves as
\beq   \label{D-2}
\bD_{(-2)} = \frac{e{\cal A}}{4\pi}\frac{\mbox{\boldmath $r$}}{r}\ ,
\eeq
where, due to the Gauss law, the monopole part of the $r$-independent
function $\cal A$ equals 1. Note that in the Maxwell case ${\cal A}\equiv 1$.
Using (\ref{D-2}) it was shown \cite{charge} that:
$$  \bH \sim r^{-1}\ ,\ \ \  \bE \sim r^{0}\ , \ \ \  \bB \sim r\ .$$
Moreover, the following theorem was proved:
\begin{TH}
Any regular solution of Born-Infeld field equations with point-like
external current satisfies:
\beq  \label{E-T}
\bE_{(1)}^T = \frac {be}{4|e|}
 \left( 3\mbox{\boldmath $a$} - 
r^{-2}(\mbox{\boldmath $a$} \mbox{\boldmath $r$})
\mbox{\boldmath $r$} \right)\ ,
\eeq
where $\bE^T$ stands for the transversal part of $\bE$ and $\ba$ denotes 
particle's acceleration.
\end{TH}
Therefore, when the particle's trajectory is {\it a priori} given, the
hyperbolicity of  (\ref{field-eqs}) implies:
\begin{TH}
The mixed (initial-boundary) value problem for the Born-Infeld equations
in ${\cal M}_\zeta$  with (\ref{E-T}) playing the role of boundary
condition on $\partial{\cal M}_\zeta$ has the unique solution.
\end{TH}

The above theorem is no longer true when we consider a particle as a dynamical
object. Choosing particle's position $\bq$ and velocity 
$\bv$ as the Cauchy data for the particle's dynamics let us observe
that despite the fact that the time derivatives $(\dot{{\bD}},\dot{{\bB}},
\dot{\bq},\dot{\bv})$ of the Cauchy data are uniquely determined
by the data themselves, the evolution of the composed system is not
uniquely determined. Indeed, $\dot{{\bD}}$ and $\dot{{\bB}}$ are given
by the field equations, $\dot{\bq}=\bv$ and $\dot{\bv}$ may
be calculated from (\ref{E-T}). Nevertheless, the initial value problem
is not well posed: keeping the same initial data, particle's trajectory 
can be modified almost at will. This is due to the fact, that now 
(\ref{E-T}) plays no longer the role of boundary condition because we
use it to as a dynamical equation to determine $\ba$. Therefore a new
boundary condition is necessary.

It was shown in \cite{charge} that this missing condition
is implied by the conservation law of the total four-momentum for
the ``particle + field'' system:
\beq   \label{conservation-4}
\dot{p}^\mu (t)=0\ ,
\eeq
where $p^\mu$ stands for the the four-momentum in any lorentzian (laboratory)
frame. Due to the field equations only 3 among 4 equations
(\ref{conservation-4}) are independent, i.e. the conservation of three-momentum
\beq   \label{conservation-3}
\dot{{\bp}} (t)=0\ ,
\eeq 
implies the energy conservation. It was shown in \cite{charge} that 
(\ref{conservation-3}) is equivalent to the following Newton-like
equation:
\beq   \label{Newton}
ma_k = \frac{|e|b}{3}{\cal A}_k\ ,
\eeq
where ${\cal A}_k$ is the dipole part of $\cal A$, i.e. DP(${\cal A}) =
:{\cal A}_k x^k/r$.
The above equation looks formally like a standard Newton equation. However,
it could not be interpreted as the Newton equation because its r.h.s. is not 
{\it a priori} given (it must be calculated from field equations).

To correctly interpret (\ref{Newton}) we have to take into account
(\ref{E-T}). Now, calculating $\ba$ in terms of $\bE^T_{(1)}$ and 
inserting into (\ref{Newton}) we obtain the following
 relation between $\bE^T_{(1)}$ and $\bD_{(-2)}$:
\beq   \label{dynamical}
   \mbox{DP} \left(   4r_0^4 (\bE^T_{(1)})^r -
\lambda_0 (\bD_{(-2)})^r  \right) =0\  ,
\eeq 
where $r_0 = \sqrt{|e|/4\pi b}$ and $\lambda_0 = e^2/6\pi m$. All constants
(like electric charge or particle's mass) enter into the dynamical condition
(\ref{dynamical})  {\it via} two characteristic scales $r_0$ and $\lambda_0$.
Obviously $r_0$ measures the nonlinearity of the Born-Infeld theory (in the
Maxwell case $r_0 \equiv 0$). The second scale $\lambda_0$
appears in any electrodynamics of charged particles, e.g. it appears in the
Lorentz-Dirac equation: 
${a}^\mu = \frac em 
F^{\mu\nu}_{ext}u_\nu + \lambda_0 (\dot{a}^\mu - a^2 u^\mu)$.
The main result of \cite{charge} consists in the following
\begin{TH}
Born-Infeld field equations supplemented by the dynamical condition
(\ref{dynamical}) define perfectly deterministic theory, i.e. initial
data for field and particle uniquely determine the entire evolution
of the system.
\end{TH}

\section{Lagrangian structure}
\label{Lagrangian}

In this section we show that the dynamical condition (\ref{dynamical})
may be derived from the mathematically well defined variational principle.
Guided by the example of Maxwell theory one could guess that such a principle
should be based on the following Lagrangian:
\beq    \label{L-standard}
L_{total} = L_{field} + L_{particle} + L_{int}\ ,
\eeq
with $L_{field}$ given by (\ref{Lag-BI}), $L_{particle}=-m\gamma^{-1}$ 
 ($\gamma^{-1} \equiv \sqrt{1-v^2}$) and $L_{int} = e A_\mu j^\mu$.
 Now, varrying
$L_{total}$ with respect to $A_\mu$ one obviously reproduces (\ref{field-eqs}). 
But the variation with respect to particle's trajectory leads to the standard
Lorentz equation
\beq   \label{Lorentz}
ma^\mu = e F^{\mu\nu} u_\nu \ . 
\eeq
Hovewer, despite the fact that $F^{\mu\nu}$ is bounded, it is not regular
at the particle's position and, therefore, the r.h.s. of (\ref{Lorentz}) is not
well defined. This was already our motivation to find the mathematically well
defined dynamical condition (\ref{dynamical}) which replaces ill defined 
equations of motion (\ref{Lorentz}).

Therefore, the variational principle based on (\ref{L-standard}) is not well
defined. To find the correct principle consider more carefully the field
dynamics with respect to an arbitrary moving observer
(see \cite{Kij-Dar} and \cite{Hamiltonian} for a general discussion). 
Consider an observer moving along an arbitrary (time-like) trajectory $\zeta$.
At each point $(t,\bq(t))\in \zeta$ let $\Sigma_t$ denotes a 3-dimensional
hyperplane orthogonal to the four-velocity vector $U=(u^\mu)$. Choose on $\Sigma_t$
any system $(x^i)$ of cartesian coordinates, such that an observer is located
at its origin. This system is not uniquely defined because on each $\Sigma_t$ 
we have still the freedom of an $O(3)$-rotation. Let {\bf L} denotes the 
boost relating the laboratory time axis $\partial/\partial y^0$ with
the co-moving observer's proper time axis $U$. Next, define the position
of the $\partial/\partial x^k$ - axis on $\Sigma_t$ by transforming the
corresponding $\partial/\partial y^k$ - axis of the laboratory frame by the
same boost. It is easy to check, that one obtains:
\beq \label{transformation}
y^0(t,\bx) &:=& t + \gamma(t)x^lv_l(t)\ ,\nonumber\\
y^k(t,\bx) &:=& q^k(t) + {\bf L}^k_{\ l}(t)x^l\ ,
\eeq
where the boost
\beq   
{\bf L}^k_{\ l} := \delta^k_{\ l} + \gamma(1+\gamma^{-1})^{-1} v^kv_l\ .
\eeq
The flat Minkowski metric tensor has in the new coordinates $(x^0=t,x^k)$
the following form: $g_{kl} = \delta_{kl}$, whereas the lapse function $N$
and the purely rotational shift vector $\bN$ read:
\beq   \label{lapse}
N &=& \frac{1}{\sqrt{-g^{00}}} = \gamma^{-1}(1 + \ba\bx)\ ,\nonumber\\
N_m &=& g_{0m} = \gamma^{-1}(\om\times\bx)_m\ ,
\eeq
where $\ba$ is the observer's acceleration vector in the co-moving frame
\beq  \label{ak}
a^i = \gamma^2 {\bf L}^i_{\ l}\dot{v}^l\ ,
\eeq
and
\beq  \label{omega}
\om = \gamma^2(1+\gamma^{-1})^{-1} \ \bv\times\dot{{\bv}}\ .
\eeq
The field equations have in the co-moving frame the following form 
(cf. a general discussion in \cite{MTW}):
\beq    \label{Maxwell}
\dot{\bD} &=& \gamma^{-1} \left[ \nabla \times (N\bH)  +
(\bN\cdot\nabla)\bD - (\bD\cdot\nabla)\bN\right] \ ,\nonumber\\
\dot{\bB} &=& \gamma^{-1} \left[- \nabla \times (N\bE)  +
(\bN\cdot\nabla)\bB - (\bB\cdot\nabla)\bN \right]    \ .
\eeq
In \cite{charge} we used a simplified (so called Fermi-Walker) frame in which
the shift vector $\bN=0$. However, such a frame is nonlocal in time
and, therefore, one can not use it to define the Hamiltonian structure
in a consistent way.
Observe, that the field evolution 
 given by (\ref{Maxwell}) is a superposition of the following 
three transformations:
\begin{itemize}
\item time-translation in the direction of the observer's four-velocity $U$,
\item boost in the direction of the observer's acceleration $\ba$,
\item purely spatial $O(3)$-rotation around the vector $\om$.
\end{itemize}
It is, therefore, obvious that this evolution is generated by the following
generator:
\beq    \label{H-field}
H_{field} := -\gamma^{-1} \left( {\cal P}^0 + \ba\bR - 
\om\bS\right)\ ,
\eeq
where ${\cal P}^0$ is the  field energy, ${\bR}$ is the static moment and
$\bS$ denotes the angular momentum. These quantities are defined in the 
observer's rest frame {\it via} a symmetric energy-momentum tensor:
\beq
T^\mu_{\ \nu} = \delta^\mu_{\ \nu}{\cal L}_{BI} 
- \frac{\partial {\cal L}_{BI}}{\partial S} F^\mu_{\ \lambda}F^\lambda_{\ \nu}
- \frac{\partial {\cal L}_{BI}}{\partial P} F^\mu_{\ \lambda}
\tilde{F}^\lambda_{\ \nu} \ . \nonumber
\eeq
The relativistic factor $\gamma^{-1}$ corresponds to the fact that
we use not the proper time along the observer's trajectory $\zeta$ but the
laboratory time. The ``$-$'' is choosen for the future convenience. 
We stress that $H_{field}$ plays the role of a Hamiltonian (with negative sign) for
any relativistic lagrangian field theory. In the case of
electrodynamics (Maxwell or general nonlinear theory) the corresponding
Hamilton equations are given by (\ref{Maxwell}).

Now, let us add to this picture a charged particle by replacing the field
energy ${\cal P}^0$ by a total "particle + field" energy ${\cal H} = m +
{\cal P}^0$. Obviously, the particle's static moment and angular momentum 
vanish in its rest frame. Define the new generator
 \beq    \label{L-H}
L_H := -\gamma^{-1} \left( {\cal H} + \ba\bR - 
\om\bS \right)\ .
\eeq
It turns out that the above generator contains all information about the 
particle's dynamics. This is due to the following
\begin{TH}
The generator $L_H$ defines the 2-nd order particle's Lagrangian, i.e. 
varrying it with respect to the particle's trajectory one recovers
the dynamical condition (\ref{dynamical}). 
\end{TH}
Proof: the Euler-Lagrange equations for a 2-nd order Lagrangian read:
\beq  \label{Euler}
\dot{\bp} = \frac{\partial L_H}{\partial \bq}\ ,
\eeq
where the momentum $\bp$ canonically conjugated to the particle's
position $\bq$ is defined by
\beq   \label{pk}
\bp :=  \frac{\partial L_H}{\partial \bv} -
\frac{d}{dt}\left(\frac{\partial L_H}{\partial \dot{\bv}} \right)
\eeq
(see \cite{Gitman} for a general discussion of  higher order Lagrangians. A 
review of a 2-nd order case may be found in \cite{Kij-Dar} and 
\cite{Hamiltonian}).
To calculate $\bp$ one needs time derivatives of $\bR$ and $\bS$. 
Using field equations (\ref{Maxwell}) and the asymtotic conditions
presented in the previous section one gets:
\beq
\dot{{\bR}} &=& \gamma^{-1} \left( {\bP} - 
\ba\times \bS -  \om\times {\bR} \right)\ ,\\
\dot{\bS} &=& \gamma^{-1} \left(  
\ba\times {\bR} - \om \times \bS \right)\ ,
\eeq
where $\bP$ denotes the field momentum in the co-moving frame.
Therefore, one obtains the following formula 
\beq
p_k = \gamma v_k {\cal H} + {\bf L}^l_{\ k}{\cal P}_l 
+ {\bf A}^l_{\ k}{\cal R}_l + {\bf B}^l_{\ k} {\cal S}_l
\eeq
with
\beq
{\bf A}^l_{\ k} &=&   \frac{d}{dt} \left( \gamma^{-1} 
\frac{\partial a^l}{\partial \dot{v}^k} \right) -
\frac{\partial (\gamma^{-1}a^l)}{\partial {v}^k} -
\gamma^{-2} \epsilon_m^{\ \ il} \left(
\frac{\partial a^m}{\partial \dot{v}^k}\omega_i +
\frac{\partial \omega^m}{\partial \dot{v}^k}a_i\right) \ ,\nonumber\\
{\bf B}^l_{\ k} &=&  - \frac{d}{dt} \left( \gamma^{-1} 
\frac{\partial \omega^l}{\partial \dot{v}^k} \right)+
\frac{\partial (\gamma^{-1}\omega^l)}{\partial {v}^k} +
\gamma^{-2} \epsilon_m^{\ \ il} \left(
\frac{\partial a^m}{\partial \dot{v}^k}a_i +
\frac{\partial \omega^m}{\partial \dot{v}^k}\omega_i\right) \ .\nonumber
\eeq
After  straightforward (but tedious) algebra one finds: ${\bf A}^l_{\ k} 
= {\bf B}^l_{\ k} =0$ and finally
\beq   \label{final-pk}
p_k = \gamma v_k {\cal H} + {\bf L}^l_{\ k}{\cal P}_l \ .
\eeq
But (\ref{final-pk}) is nothing else than a total "particle + field"
momentum in the laboratory frame. Therefore, the Euler-Lagrange equations
(\ref{Euler}) reduce to the conservation law of the total momentum (since the 
r.h.s. of (\ref{Euler}) vanishes  in our case)  and
it was proved in \cite{charge} that it is equivalent to the dynamical
condition (\ref{dynamical}). This ends the proof.

\section{Hamiltonian}
\label{Hamiltonian}

To find the corresponding Hamiltonian structure of the above theory one
has to perform the Legendre transformation to $L_H$. Let $\bpi$ denotes
the momentum canonically conjugated to particle's velocity $\bv$, i.e.
\beq   \label{pik}
\bpi := \frac{\partial L_H}{\partial \dot{{\bv}} }\ .
\eeq
Observe, that due to the fact
that $L_H$ is linear in $\dot{{\bv}}$ (see (\ref{ak}) and (\ref{omega}))
it is impossible to invert (\ref{pik}) (i.e. to calculate $\dot{{\bv}}$ in
terms of $\bpi$) and, therefore, the Legendre transformation is 
singular. It means that in the Hamiltonian framework the phase space 
of our system 
$$  {\cal P} = (\bq,\bp,\bv,\bpi;\bD,\bB)$$
 is subjected to some constraints.
To find these constraints let us apply the standard Dirac-Bergmann procedure
\cite{Dirac-Bergmann} (see also \cite{constraints}). The phase space 
${\cal P}$ is endowed with the following canonical Poisson bracket:
\beq    \label{Poisson}
\{{\cal F},{\cal G}\} = 
\frac{\partial {\cal F}}{\partial \bq} \cdot
\frac{\partial {\cal G}}{\partial \bp}  +
\frac{\partial {\cal F}}{\partial \bv} \cdot
\frac{\partial {\cal G}}{\partial  \bpi}  +
\int_{\Sigma} 
\frac{\delta {\cal F}}{\delta \bD} \cdot \nabla\times
\frac{\delta {\cal G}}{\delta \bB}\,d^3x  - \ 
({\cal F} \rightleftharpoons {\cal G})\ .
\eeq
Now, the (unreduced) Hamiltonian on ${\cal P}$ is defined by:
\beq    \label{H1}
H(\bq,\bp,\bv,\bpi;\bD,\bB) =
\bp\bv + \bpi\dot{\bv} - L_H\ .
\eeq
Therefore, the primary and secondary constraints read:
\beq    \label{primary}
\phi^{(1)}_k &:=& \pi_k - \frac{\partial L_H}{\partial \dot{v}^k}
\approx 0\ ,\\    \label{secondary}
\phi^{(2)}_k &:=& \{ \phi^{(1)}_k, H \} =
- p_k + \gamma v_k {\cal H} + {\bf L}^l_{\ k}{\cal P}_l \approx 0\ ,
\eeq
where the symbol "$\approx$" refers to the "weak equality". Observe, that
the secondary constraints $\phi^{(2)}_k = 0$ reproduce (\ref{final-pk}).
Using (\ref{primary}) and (\ref{secondary}) we get
\beq    \label{H2}
H = \bp\bv + \gamma^{-1}{\cal H} + \dot{v}^k\phi^{(1)}_k
\eeq
and $\dot{v}^k$ plays now the role of Lagrange multiplier. Let us
continue with the Dirac-Bergmann procedure and look for the tetriary 
constraints:
\beq    \label{tetriary}
\phi^{(3)}_k &:=& \{ \phi^{(2)}_k, H \} \approx 0\ ,
\eeq
with $H$ given by (\ref{H2}). One may show (following the calculations 
in \cite{charge})  that (\ref{tetriary}) implies
\beq
\dot{v}^k = \gamma^{-2}\frac{|e|b}{3m} ({\bf L}^{-1})^k_{\ l}{\cal A}^l\ ,
\eeq
which is equivalent to (\ref{Newton}). Together with (\ref{E-T}) it implies
our dynamical condition (\ref{dynamical}). Therefore, the consistency of
the theory requires that the dynamical condition has to be already present
in the definition of the physical phase space. This way the tetriary 
constraints are identically satisfied and the constraint algorithm stops
at this point. Observe, that the Lagrange multiplier $\dot{{\bv}}$ in
(\ref{H2}) is, therefore, already known which means that the constraints 
$\phi^{(1)}_k$
and  $\phi^{(2)}_k$ are of the second class (see the next section).

Now, on the
reduced phase space, i.e. ${\cal P}$ subjected to (\ref{primary})
and (\ref{secondary}) (and to the dynamical condition (\ref{dynamical}))
the reduced Hamiltonian 
\beq   \label{H-qv}
H^*(\bq,\bv;\bD,\bB) = \gamma \left( {\cal H} + \bv\bP\right)\ 
\eeq
equals to the total energy of the composed system in the laboratory frame.
In deriving (\ref{H-qv}) we chose as independent variables (in the particle's
sector) particle's position $\bq$ and velocity $\bv$ (this choice is
strongly sugested by the form of constraints equations). Obviously, this
parameterization is not unique. For example we could take as independent 
variables $\bq$ and $\bp$ as well. One could show (see 
\cite{Hamiltonian}) that 
\beq    \label{v-p}
\bv = \frac{ [\bp(\bp -\bP)] (\sqrt{{\cal H}^2 +\bp^2 - \bP^2} 
- {\cal H}) }{ [\bp(\bp-\bP)]^2 + {\cal H}^2(\bp -\bP)^2}
(\bp-\bP)
\eeq
and, therefore
\beq  \label{H-qp}
H^*(\bq,\bp;\bD,\bB) = \sqrt{{\cal H}^2 +\bp^2 
- \mbox{\boldmath ${\cal P}^2$} }\ .
\eeq
Obviously, numerically $H^*(\bq,\bv;\bD,\bB) = H^*(\bq,\bp;\bD,\bB)$. 
Observe, that for a free particle, i.e. $e=0$, ${\cal H} =m$ and 
$\bP=0$, the complicated formula (\ref{v-p}) reduces to the 
relativistic relation  between particle's velocity and momentum:
\beq
   \bv = \frac{\bp}{\sqrt{m^2 + \bp^2}}\ , \nonumber
\eeq
and formula (\ref{H-qp}) reproduces the relativistic particle's energy:
$E(\bp)=\sqrt{m^2 + \bp^2}$.

\section{Poisson bracket}
\label{bracket}
In this section we reduce the Poisson bracket (\ref{Poisson}) (defined
on ${\cal P}$) on the reduced phase space
$$   {\cal P}^* = (\bq,\bv;\bD,\bB) \ . $$
This is possible because, as one can easily check, the constraints
(\ref{primary}) and (\ref{secondary}) are of the second class:
\beq  \label{phi-phi}
\{ \phi^{(1)}_k,\phi^{(1)}_l\} &=& 0\ ,\nonumber\\
\{ \phi^{(1)}_k,\phi^{(2)}_l\} &=& -\gamma m (g_{kl} + \gamma^2v_kv_l)
\ ,\\
\{ \phi^{(2)}_k,\phi^{(2)}_l\} &=& \frac{\gamma|e|b}{3} (v_k{\cal A}_l
- v_l{\cal A}_k)\ .\nonumber
\eeq
Therefore, due to the rules of Dirac-Bergmann procedure \cite{Dirac-Bergmann}
we obtain the following formula for the reduced Poisson (or Dirac) bracket
on ${\cal P}^*$:
\beq   \label{Dirac-bracket}
\{ {\cal F},{\cal G}\}^* &=& \{ {\cal F},{\cal G}\}  +
{\bf X}^{kl} \left( \{ {\cal F},\phi^{(2)}_k\}\{\phi^{(1)}_l,{\cal G}\}
- \{ {\cal F},\phi^{(1)}_k\}\{\phi^{(2)}_l,{\cal G}\} \right) \nonumber\\
  &-&
{\bf Y}^{kl}  \{ {\cal F},\phi^{(1)}_k\}\{\phi^{(1)}_l,{\cal G}\}\ ,
\eeq
where
\beq
{\bf X}^{kl} &=&  \frac{g^{kl} - v^kv^l}{\gamma m}\ ,\\
{\bf Y}^{kl} &=& \frac{|e|b}{3m^2\gamma^3}(v^k{\cal A}^l
- v^l{\cal A}^k) \ .
\eeq  
In particular one easily shows that in the "particle's sector":
\beq
\{ q^k,q^l \}^* &=& 0\ ,\\
\{ q^k,v^l \}^* &=& {\bf X}^{kl}\ ,\\
\{ v^k,v^l \}^* &=& {\bf Y}^{kl}\ .
\eeq
 Others "commutation relations" between variables parameterizing 
${\cal P}^*$ may be easily obtained from (\ref{Dirac-bracket}).

Observe, that using (\ref{Dirac-bracket}) we are able to reproduce 
the dynamics of our theory:
\beq   \label{q-dot}
\dot{q}^k &=& \{ q^k, H^* \}^* = v^k \, \\  \label{v-dot}
\dot{v}^k &=& \{ v^k, H^* \}^* = \gamma^{-2}\frac{|e|b}{3m} 
({\bf L}^{-1})^k_{\ l}{\cal A}^l\ ,
\eeq
and in the "field sector" one easily finds that
\beq
\dot{D}^k &=& \{ D^k, H^* \}^* \, \nonumber\\
\dot{B}^k &=& \{ B^k, H^* \}^*\nonumber
\eeq
are equivalent to (\ref{Maxwell}) where  $\dot{{\bv}}$ is given
now by (\ref{v-dot}). Observe, that (\ref{v-dot}) is nothing more
than an identity because the dynamical condition (which is equivalent
to (\ref{v-dot}) together with (\ref{E-T})) is already present in the
definition of ${\cal P}^*$.
This way we have proved 
\begin{TH}
The triple $({\cal P}^*,H^*,\{\ ,\ \}^*)$ defines mathematically consistent
canonical structure of a point-like charge particle interacting
with nonlinear Born-Infeld electromagnetism.
\end{TH}
As a simple implication one can prove
\begin{TH}
Laboratory-frame Lorentz generators:
\begin{eqnarray*}
p^0 &=& \gamma ({\cal H} + v^l{\cal P}_l )\ ,\\
p_k &=& \gamma v_k {\cal H} + {\bf L}^l_{\ k}{\cal P}_l\ ,\\
r_k &=& \gamma^3 ({\bf L}^{-1})^l_{\ k} {\cal R}_l +
\gamma \epsilon_{klm} v^l {\cal S}^m + q_kp^0\ ,\\
s_k &=& \gamma^3 ({\bf L}^{-1})^l_{\ k} {\cal S}_l -
\gamma \epsilon_{klm} v^l {\cal R}^m + \epsilon_{klm}q^lp^m
\end{eqnarray*}
form with respect to $\{\ ,\ \}^*$ the Poincar\'e algebra.
\end{TH}
This shows that the canonical structure of our theory is 
perfectly consistent with the relativistic invariance.

\section{Concluding remarks}
\label{remarks}

Finally, let us make the following remarks:

1. A formalism applied in the present paper is perfectly gauge-invariant. 
There is no need to use a gauge potential to couple a particle to the field.

2. The 2-nd order particle's Lagrangian $L_H$ can not be written 
in the form as in formula (\ref{L-standard}). In particular there is no
"interaction term" in $L_H$. All information about the interaction
between a particle and fields is contained in the asymptotic conditions
for the electromagnetic field in the vicinity of particle's trajectory. 
Observe that $L_H$ serves as a Lagrangian for a particle's dynamics and
a Hamiltonian for the field dynamics. Therefore, it is a nontrivial example
of a so called Routhian function known from  analytical mechanics.

3. A total four-momentum $p^\mu$ of the composed "particle + field" system
lies always (as it should) in the future lightcone (this obvious
property does not hold in the Maxwell case \cite{KIJ}).

4. The remarkable feature of the reduced Poisson bracket is that it is 
analytical in the point $e=0$. In our opinion  this property is crucial 
for the construction of the consistent electrodynamics of point-like
objects. In the Maxwell case it is well known that all unphysical
solutions to the Lorentz-Dirac equation are nonanalytical at $e=0$ 
\cite{Rohrlich}. It turns out \cite{Hamiltonian} that this nonanalyticity
is also present in the Hamiltonian framework where we do not have
any equations of motion (there is an analog of the dynamical condition).
Namely, the Poisson bracket is nonanalytical in $e=0$.
Therefore, the analyticity of the canonical structure is a necessary 
condition for the consistency of the theory.
This point will be carefully discussed in the next paper.

5. It is clear from (\ref{phi-phi}) that in the case of purely 
electromagnetical particle (i.e. when the effective mass $m=0$)
the constraints $\phi^{(1)}_k$ and $\phi^{(2)}_k$ are no longer of a second 
class --  $\phi^{(1)}_k$ are first class. Therefore, they give rise to a
certain gauge freedom and the reduced phase space ${\cal P}^*
= (\bq,\bv;\bD,\bB)$ is not a proper space of states in this case, i.e.
the dynamics of our system can not be consistently reduced on ${\cal P}^*$.
 This means that the data $(\bq,\bv;\bD,\bB)$ does not determine
the evolution uniquely (there is a gauge freedom). The observation
that the purely electromagnetical particles do not have consistent
Hamiltonian formulation  was made long ago by Pryce 
(see section~8 of \cite{Pryce}).

6. Finally, observe that variables $(\bq,\bv;\bD,\bB)$ are highly 
noncanonical with respect to the reduced Poisson bracket. 
It would be interesting to find a canonical set.

\noindent
The results obtained in the present paper could be generalized

1. to include many charged particles,

2. to include Born-Infeld dyons, i.e. particles posessing both electric
and magnetic charges.

\noindent
Work on these points is in progress.

\section*{Acknowledgements}

 I would like to thank Prof. J. Kijowski  
for many discussions about the canonical formalism
for classical field theories, Prof. I. Bia{\l}ynicki-Birula
for pointing out the reference \cite{Pryce} and   Prof. H. R\"omer 
 for his interest in this work.  
The financial support from the Alexander von Humboldt Stiftung
is gratefully acknowledged.

\end{document}